\documentclass[preprintnumbers,
amsmath,amssymb,floatfix,10pt,prd,onecolumn,
superscriptaddress,longbibliography,nofootinbib]{revtex4-2}

\usepackage{graphicx}
\usepackage{dcolumn}
\usepackage{bm}
\usepackage[utf8]{inputenc}
\usepackage{textgreek}
\usepackage{palatino}
\usepackage{subcaption}
\usepackage{ragged2e}
\usepackage{xcolor}
\usepackage{hyperref}
\hypersetup{           
	colorlinks=true,                
	breaklinks=true,                
	urlcolor= blue,                
	linkcolor= magenta,                
	bookmarksopen=false,
	filecolor=black,
	citecolor=red,
	linkbordercolor=blue}
\usepackage{orcidlink}
\begin{document}

%\preprint{APS/123-QED}

\title{Viability Constraints on Baryogenesis in $f(R, L_{m}, T)$ Gravity}% Force line breaks with \\

\author{Kalyan Malakar\orcidlink{0009-0002-5134-1553}}
\affiliation{Department of Physics, Dibrugarh University, Dibrugarh, 786004, Assam, India}
 \affiliation{Department of Physics, Silapathar College, Dhemaji, 787059, Assam, India}
 \email{kalyanmalakar349@gmail.com}
 
 \author{Rajdeep Mazumdar\orcidlink{0009-0003-7732-875X}}
\affiliation{Department of Physics, Dibrugarh University, Dibrugarh, 786004, Assam, India}
 \email{rajdeepmazumdar377@gmail.com}

\author{Mrinnoy M Gohain\orcidlink{0000-0002-1097-2124}}
\affiliation{Department of Physics, Dibrugarh University, Dibrugarh, 786004, Assam, India}
 \affiliation{
 Department of Physics, DHSK College, Dibrugarh, 786001, Assam, India}
 \email{mrinmoygohain19@gmail.com}

\author{Kalyan Bhuyan\orcidlink{0000-0002-8896-7691}}
\affiliation{Department of Physics, Dibrugarh University, Dibrugarh, 786004, Assam, India}
\affiliation{Theoretical Physics Division, Centre for Atmospheric Studies, Dibrugarh University, Dibrugarh, 786004, Assam, India}
 \email{kalyanbhuyan@dibru.ac.in}

%\date{}% It is always \today, today,

\begin{abstract}
Our study explores gravitational baryogenesis in the context of \(f (R, L_m,T)\) gravity, where \(R\) is the Ricci scalar, \(L_m\) represents the Lagrangian density of the matter field and \(T\) stands for the metric contraction of \(T_{\mu\nu}\). We focus on models: (i) \(f (R, L_m,T)=\alpha R + \beta L_m + \gamma T\) and (ii) \(f (R, L_m,T)=\alpha R + \beta L_m^2 + \gamma T^2\), to examine the parameter constraints for a successful baryon asymmetry generation in the radiation dominated era of the cosmos under the consideration of a power-law-type cosmic expansion. The computed baryon-to-entropy ratio is found to be consistent with the measured order of the asymmetry ratio \(9.42\times10^{-11}\). In addition, the study is further extended to the generalised framework of gravitational baryogenesis, where the result shows strong agreement with the current observational data. Our findings indicate that the \(f (R, L_m,T)\) framework ensures a compatible theoretical foundation for producing the observed matter imbalance of the cosmos, thus emphasising its potential significance in early–universe cosmology.
\end{abstract}


\maketitle

\vspace{1em} 

%\tableofcontents

\section{Introduction}
One of the biggest enigmas in the present-day cosmology is why our universe is exceedingly made up of matter compared to antimatter. This puzzle is broadly referred to as baryogenesis \cite{dolgov1988cosmology}. Fundamental theories indicate that the early universe should have contained matter and antimatter in identical measure, leading to the annihilation of each other \cite{weinberg1995quantum}. However, the fact that we exist in a matter-dominated universe means that something unknown in the early era must have tipped the scales. Analysis of astronomical observations from the large-scale structures (LSS), the Cosmic Microwave Background (CMB)  \cite{bennett2003microwave}, and the predictions of Big Bang Nucleosynthesis (BBN) \cite{burles2001big} -collectively suggest a visible cosmos with matter predominance. Antimatter, if present at all, is limited to small and isolated regions, with no evidence for large-scale antimatter domains \cite{steigman1976observational,dolgov2001matter,cohen1998matter}. This observational reality implies that some physical mechanism must be actively operating to produce the detected matter imbalance in the primordial universe \cite{dine2003origin}. 
Sakharov, in 1967, classified three essential criteria that must be met to account for the minute baryonic content imbalance observed in the cosmos \cite{sakharov1998violation}:
\begin{enumerate}
\item Baryon number violation: In the earliest stages of cosmic evolution, interactions capable of violating the baryon number must have taken place. Such processes are necessary to generate a net matter surplus.
\item Non-conservation of \(\mathcal{C}\) and \(\mathcal{CP}\) symmetries: If both \(\mathcal{C}\)- \& \(\mathcal{CP}\)- were strictly conserved, any baryon-generating processes would be exactly counterbalanced, preventing the emergence of an asymmetry. Their violation is therefore a fundamental requirement.
\item Departure from Thermal Balance: While in thermal balance, forward and backward reactions progress at identical speeds, erasing any imbalance. Hence, at certain epochs, the universe must have undergone an unbalanced phase of evolution.
\end{enumerate}
 
There are numerous proposed frameworks to address the matter surplus in the cosmos. Gravitational baryogenesis provides one possible explanation for the imbalance, however, it is not the only theoretical pathway to resolve this long-standing problem. Notable alternative approaches include spontaneous baryogenesis \cite{brandenberger2003spontaneous,takahashi2004spontaneous}, electroweak baryogenesis \cite{trodden1999electroweak}, the Affleck–Dine mechanism \cite{dine2003origin,stewart1996affleck}, scenarios based on grand unified theories (GUTs) \cite{kolb1996grand}, and approaches where baryon asymmetry emerges through black hole evaporation \cite{ambrosone2022towards}. 
 
Among the various theoretical approaches, gravitational baryogenesis provides a compelling scenario, where the baryonic current, \(J^{\mu}\), interacts with the derivative of \(R\), thereby inducing \(\mathcal{CP}\)-violating effects that establishes the dominance of matter \cite{davoudiasl2004gravitational}. The coupling term that leads to the violation of \(\mathcal{CP}\) is given by:

\begin{equation}
\frac{1}{M_*^2}
\int (\partial_{\mu}R) J^{\mu}\ \sqrt{-g} d^4x \,\;,
\label{eq:coupling_equation}
\end{equation}
where, \(M_*\) corresponds to the cutoff scale in the effective field framework and \(det(g_{\mu\nu})=g\).

The notion of gravitational baryogenesis originates from the work in \cite{davoudiasl2004gravitational}, where dynamical \(\mathcal{CP}\)- violation drives the observed matter–antimatter imbalance. Based on that idea, Lambiase and Scarpetta \cite{lambiase2006baryogenesis} showed that \(f(R)\) gravity naturally accommodates this mechanism, thereby offering a potential arena for generating the required asymmetry. Subsequently, Ramos at el. \cite{ramos2017baryogenesis} addressed the same domain and successfully produced the asymmetry ratio in alignment with observations. Odintsov et al. demonstrated that both Gauss–Bonnet gravity \cite{odintsov2016gauss,Chakraborty2023Jan} and Loop Quantum Cosmology \cite{odintsov2016loop} can successfully reproduce asymmetry ratios consistent with current observational constraints. Other important gravity models include torsion-based \(f(T)\) gravity \cite{oikonomou2016f} and non-metricity inspired \(f(Q,T)\) model \cite{snehasish2020baryogenesis}. More recently, Samaddar et al. \cite{Samaddar2024May} demonstrated baryon asymmetry generation in \(f(Q,C)\) gravity. Within the \(f(R, T)\) formulation, studies have shown that the procedure of gravitational baryogenesis can yield viable asymmetry ratios consistent with observations \cite{sahoo2020gravitational,nozari2018baryogenesis,baffou2019f}. Proposed by Harko at el. \cite{harko2011f}, this hybrid gravity framework, incorporating explicit geometry-matter couplings, provides an extended formulation of GR to address matter surplus in the primordial cosmos. Under the formalism of \(f(R, L_{m})\) gravity, gravitational baryogenesis has been investigated as a possible explanation for the measured baryon asymmetry, with recent studies indicating that the model is capable of producing baryon-to-entropy ratios compatible with observational bounds in a spatially flat radiation dominated universe \cite{jaybhaye2023baryogenesis}. 

Haghani and Harko \cite{haghani2021generalizing} proposed the foundation of \(f(R, L_{m}, T)\) gravity that combine, expand and generalise two widely studied geometry–matter coupling models \(f(R, T)\) and \(f(R, L_{m})\) by promoting \(R\), \( L_{m}\) and \(T\) into a single function. In this construction, \(f(R)\), \(f(R,T)\) and \(f(R, L_{m})\) are recovered as limiting cases. Arora at el. \cite{arora2024energy} further studied the stability and energy conditions of the \(f(R, Lm, T)\) model. The cosmography of the model under consideration is discussed in \cite{mishra2025cosmography}. It must be noted that for GR, the asymmetry ratio is in direct proportion to temporal differentiation of \(R\). For a flat radiation-dominated cosmos, marked by the EoS (equation of state) parameter \(\omega=\frac{1}{3}\), \(\dot{R}\) vanishes and hence Eq. \eqref{eq:coupling_equation} yields no net baryon asymmetry. The aim of this study is to explore \(f(R, Lm, T)\) theory as a potential foundation for gravitational baryogenesis in a flat radiation dominated universe under a power-law expansion scenario, a setting that has not yet been examined for its implications on matter asymmetry.

The sections are organized in the following way. In Section \ref{sec2}, an overview of the \(f(R,L_{m},T)\) framework is presented along with derivation of the corresponding field equation. Section \ref{sec3} \& \ref{sec4} is devoted to the formulation of gravitational baryogenesis and the generalized scenario in the realm of modified \(f(R,L_{m},T)\) models, where analytical developments are carried out in detail. This section also incorporates graphical plots followed by a critical examination and interpretation of the key findings obtained. Finally, Section \ref{sec5} provides the concluding remarks and a summary of the main outcomes of the work.

\section{Description of \(f(R, L_{m}, T)\) gravity and Field Equations}
\label{sec2}
\subsection{Review on \(f(R, L_{m}, T)\) Gravity}
This paper addresses a modified theory of gravity where matter interacts non-minimally with spacetime geometry incorporated through the Lagrangian of matter-field \(L_{m}\) and the scalar contraction of \(T_{\mu\nu}\), yielding a generalized Lagrangian density articulated as \(f(R, L_{m}, T)\).  Thus, the corresponding action for the non-minimally coupled matter–geometry framework of gravity model can be expressed as \cite{haghani2021generalizing}:

\begin{equation}
S_{fRL_{m}T}=
\frac{1}{16\pi}\int \sqrt{-g}d^4x \,f(R,L_{m},T) + \int\sqrt{-g} d^4x \,L_{m},
\label{eq:action_equation}
\end{equation}

The dynamics of the cosmological equations are examined for two common alternatives for the Lagrangian density of matter field: \(L_{m}=p\) and \(L_{m}=-\rho\). For this work, our analysis is carried out using the former case, \(L_{m}=-\rho\).

Varying the action in Eq. \eqref{eq:action_equation} w.r.t. \(g^{\mu\nu}\) leads to the variation of the action functional: 

\begin{equation}
\delta S_{fRL_{m}T} = \frac{1}{16\pi}\int \sqrt{-g}d^4x \,[(f_{L} \frac{\delta L_m}{\delta g^{\mu\nu}} + f_{T} \frac{\delta T}{\delta g^{\mu\nu}} - 8\pi T_{\mu\nu} - \frac{1}{2} g_{\mu\nu} f) \delta g^{\mu\nu} + f_{R} \delta R] \, .
\label{eq:Variation_of_action}
\end{equation}

\(f_R\), \(f_L\) and \(f_T\) represents differentiation of \(f(R,L_{m},T)\) with respect to \(R\), \(L_m\) and \(T\) respectively. 

\(T_{\mu\nu}\) represents the energy-momentum tensor, having the form:  

\begin{equation}
T_{\mu\nu}=-2\frac{1}{\sqrt{g}}\frac{\delta (\sqrt{-g}L_{m})}{\delta g^{\mu\nu}} = -2\frac{\partial L_m}{\partial g^{\mu\nu}} + g_{\mu\nu} L_m.
\label{eq:energy-momentum_tensor}
\end{equation}

Setting $\delta S_{fRL_{m}T}$ equal to zero, we can deduce the field equation. Thus, the field equation takes the following form: 

\begin{equation}
R_{\mu\nu} f_R - (\nabla_{\mu} \nabla_{\nu}-g_{\mu\nu} \Box) f_R - [\frac{1}{2} (f_L + 2 f_T) + 8 \pi]T_{\mu\nu} + \frac{1}{2} [(2 f_T + f_L) L_m - f] g_{\mu\nu} - f_T \tau_{\mu\nu} =0.
\label{eq:General_field_equation}
\end{equation}

The quantity \(\tau_{\mu\nu}\) can be expressed:

\begin{equation}
\tau_{\mu\nu}=2g^{\psi\xi} \frac{\partial^2 L_m}{\partial g^{\mu\nu} \partial g^{\psi\xi}}.
\label{eq:Tau-mu_nu}
\end{equation}

Scalar contraction of Eq. \eqref{eq:General_field_equation}, yields the contracted form of the field equation:  

\begin{equation}
Rf_R + 2[(2f_T + f_L)L_m - f] - f_T \tau - T[8\pi + \frac{1}{2} (2f_T + f_L)] + 3\Box f_R = 0.
\label{eq:trace_field_equation}
\end{equation}

By using Eq. \eqref{eq:trace_field_equation}, the field equation \eqref{eq:General_field_equation} is modified into traceless representation of \(f(R,L_{m},T)\), as follows:

\begin{equation}
R_{\mu\nu} - \frac{1}{4} g_{\mu\nu} R = \frac{1}{f_R}(T_{\mu\nu} - \frac{1}{4} T g_{\mu\nu})[8\pi + \frac{1}{2}(2f_T + f_L)] + \frac{f_T}{f_R}(\tau_{\mu\nu} - \frac{1}{4} g_{\mu\nu \tau}) - \frac{1}{f_R}(\frac{1}{4} g_{\mu\nu} \Box - \nabla_{\mu} \nabla_{\nu})f_R.
\label{eq:traceless_representation_field_equation}
\end{equation}

It is worth highlighting that the Eq. \eqref{eq:traceless_representation_field_equation} transforms into the Field Equation of GR, for the special condition: \(f(R,L_{m},T)=f(R)\). Thus, it acts as limiting case of the theory.

Covariant derivative of Eq. \eqref{eq:General_field_equation} leads to the non-conservation condition of \(T_{\mu\nu}\). The equation of non-conservation is obtained as follows: 

\begin{equation}
\nabla_{\mu}(T^{\mu\nu})=\frac{1}{f_m + 8\pi} [\nabla_{\nu}(f_m L_m) - \frac{1}{2} (f_L \nabla_{\nu} L_m + f_T \nabla_{\nu} T) - A_{\nu} - T_{\mu\nu} \nabla^{\mu} f_m].
\label{eq:non_conservation_energy_momentum_tensor}
\end{equation}

where, it is defined as:

\begin{eqnarray}
A_{\nu} &= \nabla^{\mu}(\tau_{\mu\nu} f_T), 
\label{eq:A_nu} 
\qquad
f_m &= (\frac{1}{2} f_L + f_T). 
\label{eq:f_m}
\end{eqnarray}

The appearance of the non-conservative nature of Eq. \eqref{eq:non_conservation_energy_momentum_tensor} is due to the inclusion of matter fields in \(f(R, L_m, T)\). It can be easily derived from Eq. \eqref{eq:non_conservation_energy_momentum_tensor} that when both \(f_T=0\) and \(f_L=0\), the conservation condition (\(\nabla_{\mu}(T^{\mu\nu})=0\)) is recovered.

The flat FLRW ansatz that describes the large-range isotropy and homogeneity of the cosmos, is expressed as \cite{rasanen2015new}:

\begin{equation}
ds^2=a^2(t)(dx^2 + dy^2 + dz^2)-dt^2.
\label{eq:flat_FLRW_metric}
\end{equation}

The choice of Eq. \eqref{eq:flat_FLRW_metric} is strongly motivated by cosmological datasets from LSS, CMB and baryon acoustic oscillations (BAO) \cite{l2017model,jimenez2019measuring,foidl2024lambda}. These measurements collectively enhance our insights into the cosmic evolution and the present accelerated epoch. 

\subsection{Field equations}
The corresponding energy and momentum flux distribution throughout spacetime is described by the energy–momentum tensor (under the perfect fluid approximation of matter) and is given by:

\begin{equation}
T_{\mu\nu}=pg_{\mu\nu} + (p+\rho)u_{\mu}u_{\nu},
\label{eq:T_as_perfect_fluid}
\end{equation}

here, \(u^{\mu}u_{\mu}=-1\). \(\rho\), \(p\) and \(u^{\mu}\) stands for energy density, pressure and four-velocity vector of the perfect fluid, respectively. 

By using Eqs. \eqref{eq:trace_field_equation} and \eqref{eq:traceless_representation_field_equation}, the modified Friedmann Equations may be deduced for the ansatz in Eq. \eqref{eq:flat_FLRW_metric} as follows: 

\begin{eqnarray}
&& \dot{H}=\frac{1}{2f_R}(H\dot{f_R}-\ddot{f_R}) - \frac{1}{2f_R}(\rho + P)(f_m + 8\pi),
\label{eq:friedmann_equation1} \\&&
2 H^2 + \dot{H} = \frac{1}{2f_R}(\ddot{f_R} + 3H \dot{f_R}) - \frac{1}{3f_R}(2L_mf_m - f) + \frac{1}{6 f_R} (3p-\rho)(8\pi + f_m).
\label{eq:friedmann_equation2}
\end{eqnarray}

Single dot and double dots over a quantity represents the cosmic time differentiation of first and second order of the quantity, respectively.

The condition for energy balance can be achieved in \(f(R,L_m,T)\) gravity theory by combining Eqs. \eqref{eq:non_conservation_energy_momentum_tensor} and \eqref{eq:T_as_perfect_fluid}, as follows:

\begin{equation}
3H(p + \rho)+\dot{\rho}= \frac{1}{f_m + 8\pi}[\frac{1}{2} f_T (\dot{T} - 2\dot{L_m}) - (\rho + L_m)\dot{f_m}].
\label{eq:energy_conservation}
\end{equation}

Eq. \eqref{eq:energy_conservation} clearly shows that for the case: \(L_m = T =0\), the \(f(R)\) gravity energy conservation condition can be restored.

Under flat FLRW setting, the algebraic form of \(R\) takes the form:

\begin{equation}
R=6(2H^2+\dot{H}).
\label{eq:ricci_scalar}
\end{equation}

From Eq. \eqref{eq:T_as_perfect_fluid}, the trace of \(T_{\mu\nu}\) can be calculated by contraction as follows: 

\begin{equation}
T=(-1+3\omega)\rho.
\label{eq:Trace_of_Energy_momentum_tensor}
\end{equation}

here, the Equation of State (EoS) parameter, \(\omega\), is defined as: \(\omega=\frac{p}{\rho}\).

\subsection{Overview of Gravitational Baryogenesis}

Cosmological theories of present-time suggests that equivalent quantities of matter and antimatter came into existence in the primordial universe, thereby producing no net baryon asymmetry. However, analytical results from astronomical observations \cite{burles2001big,bennett2003microwave} and the non-detection of large-scale annihilation radiation \cite{cohen1998matter} strongly imply a supremacy of matter in the visible universe. This matter surplus is commonly characterized through a dimensionless quantity termed as Baryon number-to-entropy ratio (BnER) as follows \cite{malakar2025gravitational}:

\begin{equation}
\frac{\eta_B}{s}=\frac{n_B - n_{\bar{B}}}{s},
\label{eq:baryon-number-to-entropy}
\end{equation}

\(n_B\) (\(n_{\bar{B}}\)) represents particle density of baryon (anti-baryon), s stands for universe's entropy.

CMB \cite{burles2001big} and BBN \cite{bennett2003microwave} observations confirm that the accepted range of value for the asymmetry is near about: \(\frac{\eta_B}{s}\cong 9.42 \times 10^{-11}\).

The interaction term in Eq. \eqref{eq:coupling_equation} to induce \(\mathcal{CP}\)-violation is modified for \(f(R,L_m,T)\) gravity in the following manner:

\begin{equation}
\frac{1}{M_*^2}
\int d^4x \,\sqrt{-g}\; \partial_{\mu}(R + L_m +T) J^{\mu}\,.
\label{eq:modified_coupling_equation}
\end{equation}

Because the quantities \(R\), \(L_m\), and \(T\) possess different physical dimensions, the coupling constants \(\alpha\), \(\beta\), and \(\gamma\) serve as normalization factors for each term to render them dimensional homogeneity.

When the cosmic temperature drops below a threshold value, known as the decoupling temperature (\(T_D\)), the \(\mathcal{CP}\)-violating interaction dissolve, leaving behind a residual excess of matter over antimatter. Consequently, the generation of baryon asymmetry effectively takes place at (\(T=T_D\)). Beyond that, the conserved asymmetry in the universe can be formulated using the interaction term defined in Eq. \eqref{eq:modified_coupling_equation} as follows:

\begin{equation}
\frac{\eta_B}{s} \cong -\frac{15}{4\pi^2}\frac{g_b}{g_*}\frac{(\dot{R} + \dot{L_m} + \dot{T})}{M_*^2T_D}.
\label{eq:general_baryontoentropyratio}
\end{equation}

In Eq. \eqref{eq:general_baryontoentropyratio}, \(g_*\) and \(g_b\)  represent the effective nos. of degrees of freedom (d.o.f) of massless particles and total intrinsic d.o.f of the baryonic matter. Dot over \(R\), \(L_m\) and \(T\) represent cosmic time (\(t\)) differentiation of the quantities. 

The transition of the cosmos between successive equilibrium states is primarily dictated by the interplay between temperature and energy. In a thermally balanced universe, the energy density \(\rho\) exhibits a direct dependence on the temperature (\(T\)), given by \cite{piattella2018lecture}:

\begin{equation}
\rho=\frac{\pi^2}{30}g_*T^4.
\label{eq:energy_density}
\end{equation}

The expansion of the cosmos is accompanied by a gradual cooling, which in turn reduces both its temperature and energy density. As the universe cools, it undergoes a sequence of phase evolution, each associated with a distinct balanced state defined by a particular symmetry. These transitions lead to the occurrence of \(\mathcal{CP}\)-violating interactions that is vital for generating the baryon asymmetry, an outcome with significant consequences for the emergence of elementary particles and the cosmic large-scale structure \cite{sugamoto1995baryon,balaji2005dynamical,huber2023baryogenesis}.

We assume the cosmic evolution to be governed by a scale factor of power-law behaviour \cite{malakar2025f}: 

\begin{equation}
a(t)=a_0t^{\xi},
\label{eq:scalefactor}
\end{equation}
here, $\xi$ is a positive \& real constant defined as: \(\xi=\frac{2}{3(1+\omega)}\).

Adopting a generalised scale factor given by Eq. \eqref{eq:scalefactor}, offers a multifaceted and analytically manageable paradigm to investigate gravitational baryogenesis. Specific choices of \(\xi\) recover the standard cosmological epochs as special cases. Concisely, the power-law formulation provides the flexibility to examine baryon asymmetry generation over a wider category of cosmic scenarios, while retaining the generalisation for broader analysis.

Using Eq. \eqref{eq:scalefactor}, the following parameters can be expressed as: 
\begin{equation}
 H=\frac{\xi}{t}, \quad R=\frac{6 \xi (2 \xi-1)}{t^2}.
 \label{eq:H_R}
\end{equation}

\section{Gravitational Baryogenesis in \(f(R, L_{m}, T)\)}
\label{sec3}

A particular form of \(f(R,L_m,T)\) gravity needs to be adopted instead of relying on its general expression, in order to solve the Eqs. \eqref{eq:friedmann_equation1} and \eqref{eq:friedmann_equation2} analytically. In this work, we focus on models: (i) \(f (R, L_m,T)=\alpha R + \beta L_m + \gamma T\) and (ii) \(f (R, L_m,T)=\alpha R + \beta L_m^2 + \gamma T^2\) \cite{haghani2021generalizing,mishra2025cosmography,arora2024energy}. This analysis aims to investigate whether the model can account for the measured matter imbalance via a \(\mathcal{CP}\)-breaking coupling. The parameters of the model under consideration are restricted to a physically viable interval such that the theoretically calculated BnER aligns closely with the BnER inferred from astronomical observations. The methodological approach involves determining \(t_D\) in terms of \(T_D\), followed by evaluating the BnER within the constraints of the chosen model using Eq. \eqref{eq:general_baryontoentropyratio}.

\subsection{Model I}

In Model I, the adopted \(f(R, L_{m}, T)\) model is expressed as follows \cite{haghani2021generalizing,mishra2025cosmography}:
\begin{equation}
f(R, L_{m}, T)=\alpha R + \beta L_m + \gamma T,
\label{eq:f_form}
\end{equation}
here, \(\alpha\), \(\beta\) and \(\gamma\) are free parameters of the model.

The energy density (\(\rho\)) expression can be deduced by using Eqs. \eqref{eq:friedmann_equation1}, \eqref{eq:ricci_scalar}, \eqref{eq:Trace_of_Energy_momentum_tensor}, \eqref{eq:scalefactor}, \eqref{eq:H_R} and \eqref{eq:f_form} as follows:

\begin{equation}
\rho=\frac{2 \alpha  \xi^2}{t^4 (-1-\omega ) \left(\frac{\beta }{2}+\gamma +8 \pi \right)}
\label{eq:energy_density_derived}
\end{equation}

By equating the Eqs. \eqref{eq:energy_density} and \eqref{eq:energy_density_derived}, a relation can be reformulated between decoupling time \((t_D)\) and decoupling temperature \((T_D)\), having the following form:

\begin{equation}
t_D = -\frac{2^{3/4} \times \sqrt[4]{15} \times \sqrt[4]{\alpha } \times \sqrt{\xi}}{\sqrt{\pi } \times \sqrt[4]{-g_* \times T_D^4 (\omega +1) (\beta +2 \gamma +16 \pi )}}.
\label{eq:tD_for_GB}
\end{equation}

The expression for BnER can be obtained by substituting Eqs. \eqref{eq:ricci_scalar}, \eqref{eq:Trace_of_Energy_momentum_tensor}, \eqref{eq:energy_density_derived} and \eqref{eq:tD_for_GB} into the Eq. \eqref{eq:general_baryontoentropyratio}, which is given below: 

\begin{equation}
\frac{\eta_B}{s} = \frac{g_b \sqrt[4]{g_* (\omega +1) (-\beta - 2 \gamma - 16 \pi ) T_D^4} \left(45 \sqrt{2} (1-2 \xi) \sqrt{g_* (\omega +1) (-\beta - 2 \gamma - 16 \pi) T_D^4}+2 \pi  \sqrt{15} \sqrt{\alpha } g_* (2-3 \omega ) T_D^4\right)}{4 \times 30^{3/4} \sqrt{\pi } \alpha ^{3/4} g_* M_*^2 \sqrt{\xi} T_D}.
\label{eq:ratio_expsn_GB}
\end{equation}

As highlighted in \cite{davoudiasl2004gravitational}, the interaction terms of the form given in Eq. \eqref{eq:coupling_equation} naturally arise within the framework of low-energy effective field theory originating from QFT in curved spacetime, given that the cut-off scale \((M_*)\) is close to the reduced Planck mass, \(M_*=\frac{m_{Pl}}{\sqrt{8\pi}}\), where \((M_p)\) denotes the Planck mass \cite{sahoo2020gravitational,nozari2018baryogenesis}. The model successfully generates the required matter imbalance ratio if \(T_D \leq M_I\), \(M_I\) is the inflationary scale. The inflation scale is estimated to be \(M_I \simeq 2\times10^{16} GeV\), inferred from the gravitational-wave signatures reported by LIGO \cite{oikonomou2016f,lambiase2006baryogenesis,ramos2017baryogenesis}. For the evaluation of Eq. \eqref{eq:ratio_expsn_GB}, the relevant parameter values are taken as: \(M_*=2\times10^{18} GeV\), \(T_D=2\times10^{9} GeV\), \(g_*=106\) and \(g_b \simeq 1\) \cite{lambiase2006baryogenesis,nozari2018baryogenesis,ramos2017baryogenesis,Boulkaboul2023May,Pereira2025Jul,Pereira2024Sep}.

\begin{figure}[htbp]
\centering
  \includegraphics[width=0.90\linewidth]{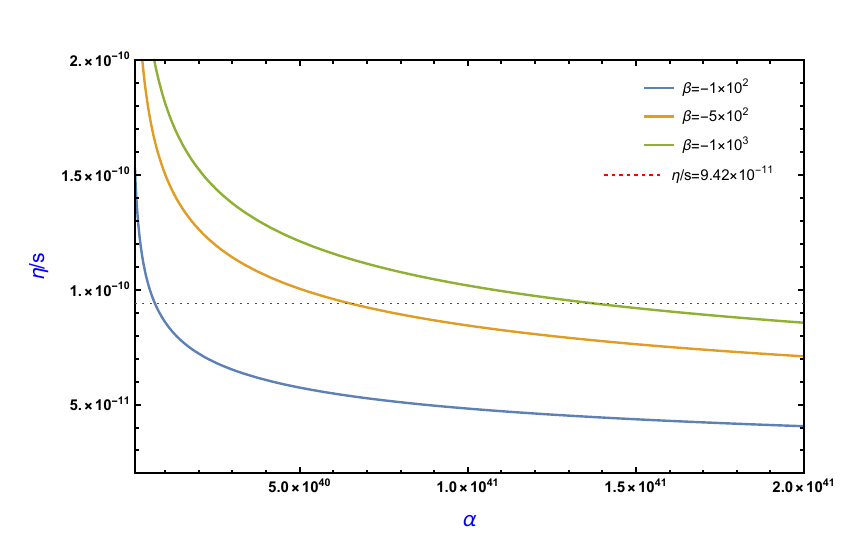}
  \caption{\justifying Plot of \(\frac{\eta_B}{s}\) versus \(\alpha\) for the \(f(R,L_m,T)\) model I under consideration in radiation era \((\omega=\frac{1}{3})\) for three different \(\beta\) choices.}
  \label{fig:Figure2}
\end{figure}

\subsubsection{Radiation-dominated era (\(\omega=\frac{1}{3}\))}

For radiation-dominated epoch, Eq. \eqref{eq:ratio_expsn_GB} gets simplified as follows:

\begin{equation}
\frac{\eta_B}{s}\;\Big|_{\omega =\frac{1}{3}} =-\frac{9.64\times 10^{-49} (-\beta -2 \gamma -50.24)^{3/4} \left(3.39\times 10^{47} \times \sqrt{\alpha(-\beta -2 \gamma -50.24)}+1.41\times 10^{14}\right)}{\alpha ^{3/4} (\beta +2 \gamma +50.24)}.
\label{eq:BnER_for_GB_radiation}
\end{equation}

Fig. \ref{fig:Figure2} illustrates the variation of BnER \((\frac{\eta_B}{s})\) versus the model parameter \(\alpha\), considering three representative choices of \(\beta=-1\times 10^{2}\), \(\beta=-5\times 10^{2}\) and \(\beta=-1\times 10^{3}\) for Model I. Throughout this analysis, the parameter \(\gamma\) is kept fixed at \(\gamma=1\). The dashed red line in the figure indicates the observationally inferred value of baryon asymmetry, \(\frac{\eta_B}{s}\). The plot depicted in Fig. \ref{fig:Figure2}, suggests that the theoretically calculated asymmetry ratio for the discussed model is consistent with observational constraints.

\subsection{Model II}

For this case, we employ the \(f(R, L_{m}, T)\) model formulated as follows \cite{haghani2021generalizing,mishra2025cosmography}:

\begin{equation}
f(R, L_{m}, T)=\alpha R + \beta L_m^2 + \gamma T^2,
\label{eq:f_model2}
\end{equation}
here, \(\alpha\), \(\beta\) and \(\gamma\) are free parameters of the model.

The energy density for Model II can be deduced in a similar manner as follows:

\begin{equation}
\rho=-\frac{2 \alpha  \xi ^2}{\sqrt{2} \sqrt{t^4 (\omega +1) \left(\alpha  \xi ^2 (\beta -6 \gamma  \omega +2 \gamma )+8 \pi ^2 t^4 (\omega +1)\right)}+4 \pi  t^4 (\omega +1)}.
\label{eq:energy_density-model2}
\end{equation}

The expression for decoupling time \(t_D\) is obtained as:

\begin{equation}
t_D=\frac{\sqrt{15} \left(\frac{2}{\pi }\right)^{3/4} \sqrt[4]{\alpha } \sqrt{\xi }}{\sqrt[4]{g_* T_D^4 (\omega +1) \left(\pi  g_* T_D^4 (\beta -6 \gamma  \omega +2 \gamma )-240\right)}}.
\label{eq:tD_model2}
\end{equation}

Hence, the final expression of BnER for Model II is obtained as: 

\begin{equation}
\frac{\eta_B}{s}=\frac{15 g_b (J (2 - 3 \omega) - K)}{4 \pi ^2 g_* M_*^2 T_D},
\label{eq:ratio_model2}
\end{equation}

where, 

\begin{align*}
J &=\frac{8 \alpha  \left(C+\frac{\pi ^{3/2} (D+G)}{40 \sqrt{F}}\right)}{9 (\omega +1)^2 \left(\frac{3200 \alpha }{\pi ^2 B g_* T_D^4 (\omega +1)^2}+\frac{40 \sqrt{F}}{\pi ^{3/2}}\right)^2},
\hspace{5cm} K=\frac{\pi ^{9/4} \left(B g_* T_D^4 (\omega +1)\right)^{3/4} \left(1-\frac{4}{3 (\omega +1)}\right)}{5\ 2^{3/4} \sqrt{5} \alpha ^{3/4} \sqrt{\frac{1}{\omega +1}}},\\
G &=\frac{160\ 2^{3/4} \sqrt{5} \alpha ^{3/4} \sqrt{\frac{1}{\omega +1}} \left(\frac{4 \alpha  A}{9 (\omega +1)^2}+\frac{6400 \alpha }{\pi  B g_* T_D^4 (\omega +1)^2}\right)}{\pi ^{9/4} \left(B g_* T_D^4 (\omega +1)\right)^{3/4}},
\hspace{2.8cm} F=\frac{\alpha  \left(\frac{4 \alpha  A}{9 (\omega +1)^2}+\frac{6400 \alpha }{\pi  B g_* T_D^4 (\omega +1)^2}\right)}{B g_* T_D^4 (\omega +1)^2},\\
D &=\frac{1024000\ 2^{3/4} \sqrt{5} \alpha ^{7/4} \left(\frac{1}{\omega +1}\right)^{3/2}}{\pi ^{13/4} \left(B g_* T_D^4 (\omega +1)\right)^{7/4}},
\hspace{5.8cm} C=\frac{640\ 2^{3/4} \sqrt{5} \alpha ^{3/4} \sqrt{\frac{1}{\omega +1}}}{\pi ^{5/4} \left(B g_* T_D^4 (\omega +1)\right)^{3/4}},\\
B &=\left(\pi  A g_* T_D^4-240\right),
\hspace{8cm} A=\left(\beta -6 \gamma  \omega +2 \gamma\right).
\end{align*}

For Eq. \eqref{eq:ratio_model2}, the set of constant parameters is taken to be identical to that used in Model I. 

\begin{figure}[htbp]
\centering
  \includegraphics[width=0.90\linewidth]{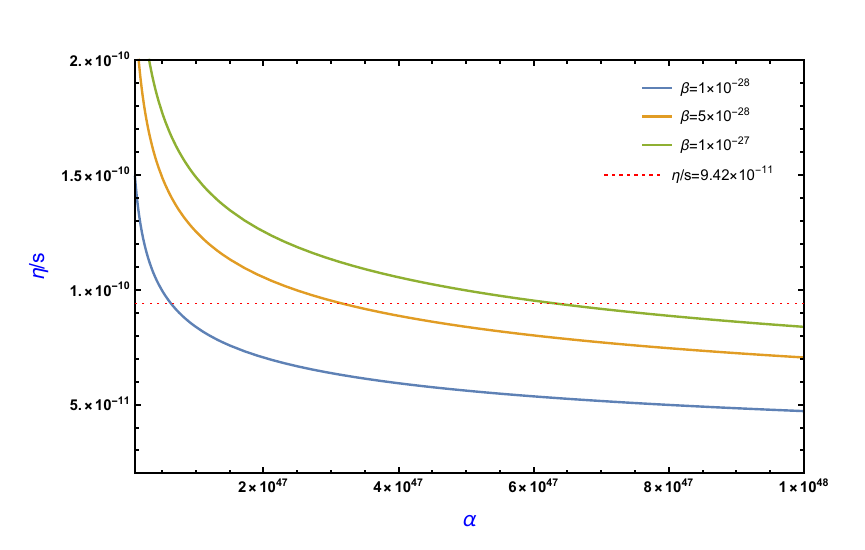}
  \caption{\justifying Plot of \(\frac{\eta_B}{s}\) versus \(\alpha\) for the \(f(R,L_m,T)\) model II under consideration in radiation era \((\omega=\frac{1}{3})\) for three different \(\beta\) choices.}
  \label{fig:Figure3}
\end{figure}

\subsubsection{Radiation-dominated era (\(\omega=\frac{1}{3}\))}

For radiation-dominated regime, BnER in Eq. \eqref{eq:ratio_model2} reduces to the following simplified expression:

\begin{equation}
\frac{\eta_B}{s}\;\Big|_{\omega =\frac{1}{3}} =\frac{\pi ^{3/4} \alpha ^{3/4} g_b \sqrt[4]{g_* T_D^4 \left(\pi  \beta  g_* T_D^4-240\right)}}{2 \sqrt{5} 6^{3/4} g_* M_*^2 T_D \sqrt{\frac{\alpha ^2 \left(\pi  \beta  g_* T_D^4-120\right)^2}{g_*^2 T_D^8 \left(\pi  \beta  g_* T_D^4-240\right)^2}}}.
\label{eq:BnER_for_GB_radiation_model2}
\end{equation}

Fig. \ref{fig:Figure3} illustrates the variation of the BnER \((\frac{\eta_B}{s})\) against the parameter \(\alpha\) for distinct \(\beta\) parameter values of \(\beta=1\times 10^{-28}\), \(\beta=5\times 10^{-28}\) and \(\beta=1\times 10^{-27}\) for Model II. In this case, the parameter \(\gamma\) is fixed at \(\gamma=1\). A comparison between the theoretical curves and BnER inferred from cosmological data, represented by the horizontal red dashed line in fig. \ref{fig:Figure3}, suggests close agreement with observations.

\section{Generalized Gravitational Baryogenesis in \(f(R, L_{m}, T)\)}
\label{sec4}

Here, we extend into the generalized case by considering a more comprehensive \(\mathcal{CP}\)-violating interaction for the chosen gravity models. The interaction term in gravitational baryogenesis is directly linked to (\(R + L_m + T\)), while, the generalized gravitational baryogenesis formulation introduces a coupling term proportional to the function itself, thereby enriching the dynamics of \(\mathcal{CP}\)-violation. The corresponding coupling term in this context in this context can thus be expressed as \cite{nozari2018baryogenesis}:

\begin{equation}
\frac{1}{M_*^2}
\int d^4x \,\sqrt{-g}\; \partial_{\mu}f(R,L_m,T) J^{\mu}\,.
\label{eq:modified_coupling_equation_GGB}
\end{equation}

From Eq. \eqref{eq:modified_coupling_equation_GGB}, the general \(\frac{\eta_B}{s}\) expression of generalized gravitational baryogenesis takes the form as:

\begin{equation}
\frac{\eta_B}{s} \cong -\frac{15}{4\pi^2}\frac{g_b}{g_*}\frac{(\dot{R} f_R + \dot{L_m} f_{L_m} + \dot{T} f_T)}{M_*^2T_D}.
\label{eq:general_baryontoentropyratio_GGB}
\end{equation}

\subsection{Model I}

By inserting Eqs. \eqref{eq:ricci_scalar},\eqref{eq:Trace_of_Energy_momentum_tensor}, \eqref{eq:f_form}, \eqref{eq:scalefactor} and \eqref{eq:tD_for_GB} into Eq. \eqref{eq:general_baryontoentropyratio_GGB}, the expression for \(\frac{\eta_B}{s}\) can be derive as follows: 

\begin{equation}
\begin{split}
\frac{\eta_B}{s}\;\Big|_{GGB} &= \left(45 \times \sqrt{2 \alpha} (2 \xi -1) \sqrt{g_* (\omega +1) (-\beta - 2 \gamma - 16 \pi ) T_D^4}-2 \sqrt{15} \pi  g_* T_D^4 (\beta -3 \gamma  \omega +\gamma )\right) \\ 
&\quad \times \frac{g_b \sqrt[4]{g_* (\omega +1) (-\beta - 2 \gamma - 16 \pi) T_D^4}}{4 \times 30^{3/4} \sqrt{\pi } \sqrt[4]{\alpha } g_* M_*^2 \sqrt{\xi } T_D}.
\label{eq:ratio_for_GGB}
\end{split}
\end{equation}

To analyse the generalized case of gravitational baryogenesis for \(f(R,L_m,T)\) model, the parameters having fixed magnitudes in Eq. \eqref{eq:ratio_for_GGB} are: \(g_*=106\), \(g_b \simeq 1\), \(M_*=2\times10^{18} GeV\), and the decoupling temperature is chosen as \(T_D=2\times10^{9} GeV\).

\begin{figure}[htbp]
\centering
  \includegraphics[width=0.90\linewidth]{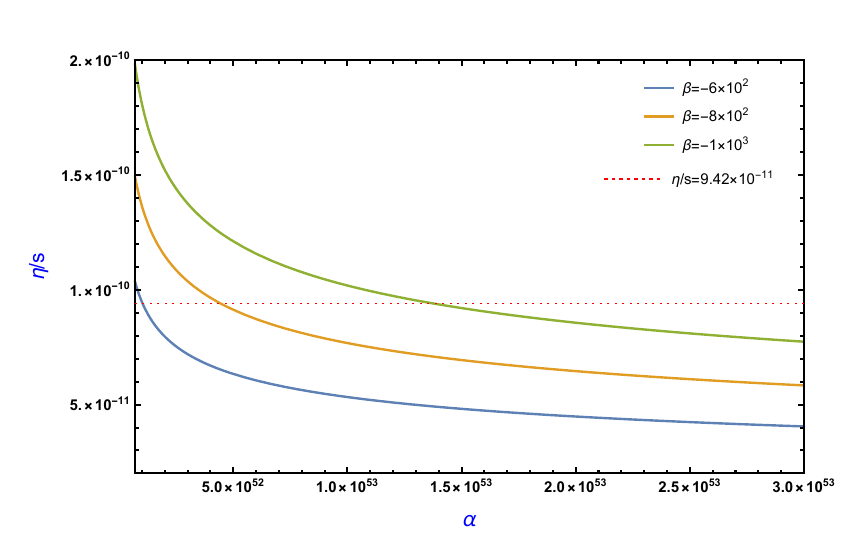}
  \caption{\justifying Plot of \(\frac{\eta_B}{s}\) versus \(\alpha\) (generalized case) for the \(f(R,L_m,T)\) model I under consideration in radiation era \((\omega=\frac{1}{3})\) for three different \(\beta\) choices.}
  \label{fig:Figure6}
\end{figure}

\subsubsection{Radiation-dominated era (\(\omega=\frac{1}{3}\))}

For radiation dominance, the expression for the asymmetry ratio given in Eq. \eqref{eq:ratio_for_GGB} takes the form:

\begin{equation}
\begin{split}
\frac{\eta_B}{s}\;\Big|_{GGB, \omega=\frac{1}{3}} &= \frac{(2.27\times 10^{-15}+9.03\times 10^{-17} \gamma )\gamma }{\sqrt[4]{\alpha } (-\beta -2 \gamma -50.24)^{3/4}}-\frac{0.33 \beta \times  (-\beta -2 \gamma -50.24)^{1/4}}{\sqrt[4]{\alpha }}-\frac{1.36\times 10^{-34} \times \sqrt[4]{\alpha }}{(-\beta -2 \gamma -50.24)^{1/4}}.
\label{eq:BnER_for_GB_radiation}
\end{split}
\end{equation}

Fig. \ref{fig:Figure6} displays the behaviour of the generalized BnER \((\frac{\eta_B}{s})_{ggb}\) in terms of \(\alpha\) for distinct choices of \(\beta\), namely \(\beta=-6\times 10^{2}\), \(\beta=-8\times 10^{2}\) and \(\beta=-1\times 10^{3}\), while keeping \(\gamma\) fixed at unity. The horizontal red dashed line in the figure corresponds to the baryon asymmetry derived from astronomical data. Theoretical estimates, in comparison with observational constraints, indicate a strong consistency between the model outcomes and empirical data.

\subsection{Model II}

For Model II, the expression of BnER is obtained in the following form: 

\begin{equation}
\frac{\eta_B}{s}\;\Big|_{GGB} = \frac{15 g_b \left(\beta  J+\gamma  (J) (3 \omega -1)^2 - K\right)}{4 \pi ^2 g_* M_*^2 T_D},
\label{eq:ratio_for_GGB_model2}
\end{equation}

where, 

\begin{align*}
J &=\frac{128 \alpha ^2 \left(C+\frac{\pi ^{3/2} (D+G)}{40 \sqrt{F}}\right)}{81 (\omega +1)^4 \left(\frac{3200 \alpha }{\pi ^2 B g_* T_D^4 (\omega +1)^2}+\frac{40 \sqrt{F}}{\pi ^{3/2}}\right)^3},
\hspace{3.6cm} K=\frac{\pi ^{9/4} \sqrt[4]{\alpha } \left(1-\frac{4}{3 (\omega +1)}\right) \left(B g_* T_D^4 (\omega +1)\right)^{3/4}}{5\ 2^{3/4} \sqrt{5} \sqrt{\frac{1}{\omega +1}}},\\
G &=\frac{160\ 2^{3/4} \sqrt{5} \alpha ^{3/4} \sqrt{\frac{1}{\omega +1}} \left(\frac{4 \alpha  A}{9 (\omega +1)^2}+\frac{6400 \alpha }{\pi  B g_* T_D^4 (\omega +1)^2}\right)}{\pi ^{9/4} \left(B g_* T_D^4 (\omega +1)\right)^{3/4}},
\hspace{1.8cm} F=\frac{\alpha  \left(\frac{4 \alpha  A}{9 (\omega +1)^2}+\frac{6400 \alpha }{\pi  B g_* T_D^4 (\omega +1)^2}\right)}{B g_* T_D^4 (\omega +1)^2},\\
D &=\frac{1024000\ 2^{3/4} \sqrt{5} \alpha ^{7/4} \left(\frac{1}{\omega +1}\right)^{3/2}}{\pi ^{13/4} \left(B g_* T_D^4 (\omega +1)\right)^{7/4}},
\hspace{4.8cm} C=\frac{640\ 2^{3/4} \sqrt{5} \alpha ^{3/4} \sqrt{\frac{1}{\omega +1}}}{\pi ^{5/4} \left(B g_* T_D^4 (\omega +1)\right)^{3/4}},\\
B &=\left(\pi  A g_* T_D^4-240\right),
\hspace{7cm} A=\left(\beta -6 \gamma  \omega +2 \gamma\right).
\end{align*}

The fixed parameters utilized in Eq. \eqref{eq:ratio_for_GGB_model2} follow the identical numerical values as those employed in earlier cases.

\begin{figure}[htbp]
\centering
  \includegraphics[width=0.90\linewidth]{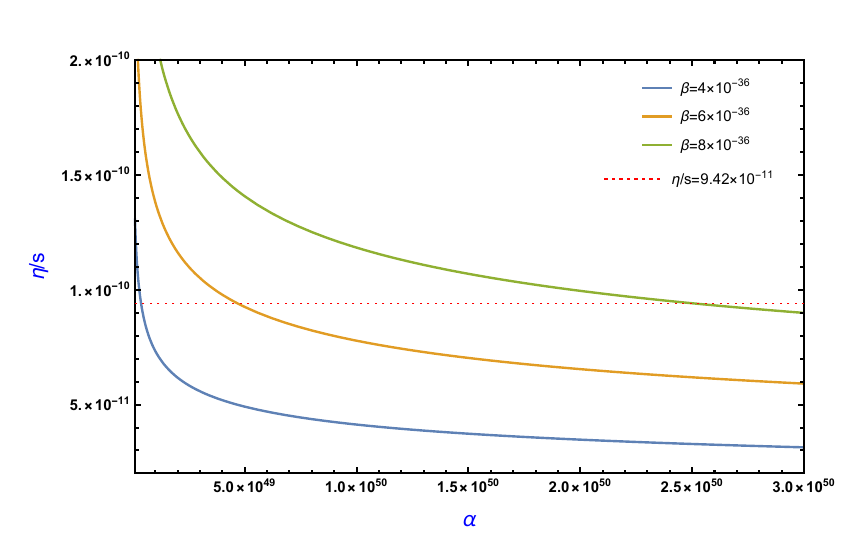}
  \caption{\justifying Plot of \(\frac{\eta_B}{s}\) versus \(\alpha\) (generalized case) for the \(f(R,L_m,T)\) model II under consideration in radiation era \((\omega=\frac{1}{3})\) for three different \(\beta\) choices.}
  \label{fig:Figure7}
\end{figure}

\subsubsection{Radiation-dominated era (\(\omega=\frac{1}{3}\))}

For radiation dominant regime, the expression for the BnER obtained from Eq. \eqref{eq:ratio_for_GGB_model2} simplifies as follows

\begin{equation}
\frac{\eta_B}{s}\;\Big|_{GGB, \omega=\frac{1}{3}} = \frac{\pi ^{11/4} \alpha ^{11/4} \beta  g_b P^{5/4} \left(28800 \alpha +\pi  \alpha  \beta  P+240 P \sqrt{Q}\right)}{30 \sqrt{5} 6^{3/4} g_* M_*^2 \sqrt{Q} T_D \left(120 \alpha +P \sqrt{Q}\right)^3},
\end{equation}

where, 

\begin{align*}
P &= \Big(g_* T_D^4 \left(\pi  \beta  g_* T_D^4-240\right)\Big),
\hspace{4cm} Q=\left(\frac{\alpha ^2 \left(\pi  \beta  g_* T_D^4-120\right)^2}{g_*^2 T_D^8 \left(\pi  \beta  g_* T_D^4-240\right)^2}\right).
\end{align*}

Fig. \ref{fig:Figure7} shows how the imbalance ratio \((\frac{\eta_B}{s})_{ggb}\) evolves with the free parameter \(\alpha\). The outcome is sensitive to the values of the \(\beta\) parameter while \(\gamma\) is fixed at unity. In fig. \ref{fig:Figure7}, three specific \(\beta\) values are taken, \(\beta=4\times 10^{-36}\), \(\beta=6\times 10^{-36}\) and \(\beta=8\times 10^{-36}\). The horizontal dashed red line in the figure marks the baryon asymmetry deduced from cosmological observations, serving as a reference point. By comparing the plotted graphs with this observational indicator, one finds strong compatibility between theoretical estimates and empirical evidence.

\section{Conclusion}
\label{sec5}

A central unresolved conundrum in cosmology is the detected excess of matter in the cosmos \cite{dine2003origin}. Among the proposed mechanisms to resolve this asymmetry is gravitational baryogenesis, which provides a possible explanation for the baryon–antibaryon imbalance \cite{davoudiasl2004gravitational,Oikonomou2015Dec}. This study investigates the phenomenon within the \(f(R,L_m,T)\) paradigm. Earlier studies provide strong motivation for this kind of exploration. For instance, the \(f(R,L_m,T)\) model satisfies energy criteria and admits stability for cosmological expansion of power-law form \cite{arora2024energy}. Furthermore, author in \cite{mishra2025cosmography} suggested that \(f(R,L_m,T)\) model presents a viable setting to address broader challenges in cosmology.

The creation of matter imbalance in the cosmos can be attributed to charge-parity (\(\mathcal{CP}\))-violating interactions, involving terms like \(\partial_{\mu}(R+L_m+T)\) and \(\partial_{\mu} f(R,L_m,T)\), which acts as key operators for this mechanism. Our analysis produced BnER ratios \(\frac{\eta_B}{s}\) in agreement with observationally established bounds, \(\frac{\eta}{s} \cong 9.42 \times 10^{-11}\). This outcome demonstrates that the \(f(R,L_m,T)\) model provides an effective explanation for the detected imbalance. Furthermore, generalized gravitational baryogenesis is also examined by modifying the interaction term to introduce extra degrees of freedom. The results show compatibility with observational constraints.

The main results of our work are outlined below:

\subsection{Model I}

\begin{itemize}

\item{GB in Model I:} 

The expression for BnER for model I is given by \eqref{eq:ratio_expsn_GB}. In this work, by restricting the asymmetry ratio $\frac{\eta}{s}$ to the observationally acceptable interval: $ 2 \times 10^{-10} \geq \eta/s \geq 7 \times 10^{-11} $, we derive a viable parameter space. By setting \(\gamma=1\) and confining \(\beta\) to the interval \(-10^{3} \; \leq \; \beta \; \leq \; -10^{2},\) the viable range of \(\alpha\) is found to be $3.5 \times 10^{38} \leq \alpha \leq 4.25 \times 10^{41}.$

\item{GGB in Model I:}

The expression for BnER within \(f(R,L_m,T)\) gravity model I under the generalized case is given by Eq. \eqref{eq:ratio_for_GGB}. To ensure that the predicted asymmetry parameter \(\eta/s\) lies within the observationally acceptable interval \( 2 \times 10^{-10} \geq \eta/s \geq 7 \times 10^{-11} \), by fixing \(\gamma = 1\) and confining \(\beta\) to the interval \(-1 \times 10^{3} \leq \beta \leq -1 \times 10^{2}\), the variable \(\alpha\) should reside in the range \(3.5 \times 10^{46} \leq \alpha \leq 4.5 \times 10^{53}\) to produce the observed baryon asymmetry.

\end{itemize}

\subsection{Model II}

\begin{itemize}

\item{GB in Model II:}

The analytical form of BnER corresponding to Model II is presented by Eq. \eqref{eq:ratio_model2}. In this analysis, the asymmetry ratio \(\eta/s\) is constrained within the observationally supported range \( 2 \times 10^{-10} \geq \eta/s \geq 7 \times 10^{-11} \), ensuring compatibility with cosmological bounds. Under the choice \(\gamma = 1\) and restricting the parameter \(\beta\) to the interval \(1 \times 10^{-32} \leq \beta \leq 1 \times 10^{-27}\), the corresponding viable range of \(\alpha\) is obtained as \(5 \times 10^{44} \leq \alpha \leq 2 \times 10^{48}\).

\item{GGB in Model II:}

The mathematical form of BnER for the gravity Model II under the generalized scenario is described by Eq. \eqref{eq:ratio_for_GGB_model2}. In order to maintain consistency with the observationally permitted range of the asymmetry parameter, \(7 \times 10^{-11} \leq \eta/s \leq 2 \times 10^{-10}\), the parameters are appropriately constrained. By setting \(\gamma = 1\) and restricting \(\beta\) within \(1 \times 10^{-36} \leq \beta \leq 1 \times 10^{-34}\), the parameter \(\alpha\) is found to lie in the range \(1 \times 10^{46} \leq \alpha \leq 4.25 \times 10^{56}\), which successfully yields the observed matter asymmetry.

\end{itemize}

Comparative analysis of the models reveals a systematic distinction in their coupling behavior. Both models operate in a regime of stronger curvature coupling (\(\alpha\)) of order \(10^{38} - 10^{56}\), showing significant departures from GR, where \(\alpha = \frac{1}{2}\). Linear Model I requires a strong and negative matter coupling (\(\beta \sim -10^{2} \text{ to } -10^{3}\)), which presents a more natural parameter setting. In contrast, non-linear Model II requires an extremely weak but positive coupling (\(\beta \sim 10^{-27} \text{ to } 10^{-36}\)). As a consequence, Model I emerges as the more theoretically plausible framework for generating the observed baryon asymmetry.

Based on the obtained results, it can be implied that \(f(R,L_m,T)\) model offers a viable setting for realizing gravitational baryogenesis, under favoured model parameter bounds. This work can be extended to study different models of \(f(R,L_m,T)\) theory under some other feasible forms of scale factor.

\bibliography{references}

\end{document}